# The hidden dimension in nanophotonics design: understanding

**Philippe Lalanne[1] and Owen D. Miller[2]**

[1] LP2N, CNRS, IOGS, Université Bordeaux, Talence, France
[2] Department of Applied Physics and Energy Sciences Institute, Yale University, New Haven, CT, USA

E-mail: philippe.lalanne@institutoptique.fr , owen.miller@yale.edu

**Abstract**: Space, time, and additional dimensions spawn remarkable complexity in optics. We encourage pairing black-box simulation and design tools with a complementary tool: understanding.

## Status

In many engineering and scientific domains, e.g. fluid dynamics, mechanics, or photonics (our focus here), the discretization of the governing differential equations results in billions of unknowns for accuracy, particularly in industrial settings, where complex, large-scale systems are commonplace. The inverse problem—designing the optimal structure—is even more challenging (See Section 5.3). Further complexity enters in “multi-physics” scenarios in which different governing equations (wave equation, heat equation, etc.) are coupled.

Despite this apparent complexity, many systems of practical interest exhibit behaviour that depends on only a small number of latent (reduced-order) parameters. Such parameters can often be uncovered through simplified models, which temporarily suppress fine-scale details and enable the formulation of more universal physical laws. A well-known example in nanophotonics is coupled mode theory [1], which assumes that a small number of modes, connected through fitted coupling coefficients, can adequately capture the system’s behaviour.

At the same time, advances in computing power, hardware architectures, and numerical algorithms have driven an increasing reliance on large-scale full-wave simulations, typically based on finite-difference or finite-element methods. In parallel, inverse problems are now frequently addressed using adjoint-based optimization or neural-network-driven approaches.

These developments give rise to a paradox in modern computational science. Although simplified, interpretable representations of physical systems are often available, there is a growing shift toward “black-box” methods and what might be termed blind simulation. This shift is likely to accelerate in the near future.

This Section does not argue against the value of full-wave simulation—on the contrary, its potential is undeniable. Rather, it raises a fundamental question: what is lost in the transition toward increasingly opaque computational methods? And, if such losses are significant, how might they be mitigated? We advocate for complementing black-box simulation and design with an equally important objective: understanding.

## Challenges: The opacity of modern computational design

In recent years, an increasing number of high-impact studies have reported near-optimal designs obtained through high-dimensional optimization driven by opaque, black-box neural networks—often referred to as foundation models—coupled with equally opaque simulation tools. This emerging approach may be described as a *black-squared* ($b^2$) paradigm. When pushed to the geometric extreme, yielding fully freeform structures (Figure 1), these designs often exhibit intricate and mysterious patterns [2]. Such structures appear to lie far beyond human intuition, fostering the impression that machines have surpassed human designers.

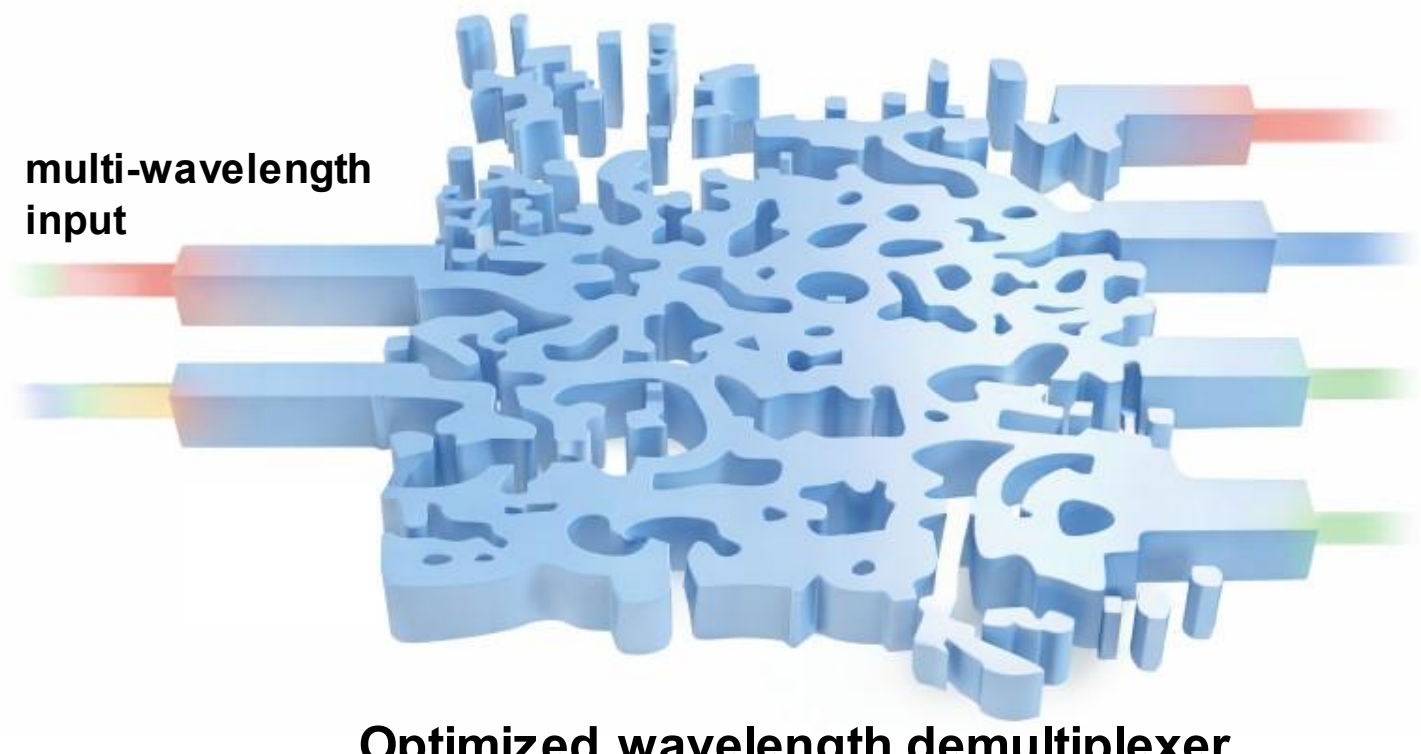

**Figure 1.** AI-driven freeform optimization in nanophotonics often yields complex *Swiss-cheese* patterns—powerful, if mysterious. Comprehension lags behind: Are the design tolerant to imperfection? Governed by multiple scattering? How close are they from upper bound physical limits, and how does that limit scale with footprint? (Figure inspired by [2]).

In a deep neural network, the layers are designed to transform raw input data (such as images, signals) with millions of variables into progressively more abstract and high-level features. Each layer operates on the output of the previous layer, which means that the deeper layers learn increasingly complex representations of the input data. This hierarchical learning process mimics human perception—starting with basic features and moving to more abstract representations.

Deep networks are often said to learn latent representations of the complex world. These latent representations are the results of non-linear combinations of features or interactions between the many design variables, which might not be easily discovered by human experts, though they may hold key insights into how the design works. However, these latent representations are not directly interpretable. The neural network does not "tell us" what the latent information represents, nor does it necessarily align with physical models that we can understand..

One parallel might be the “shut up and calculate” paradigm in quantum physics—should we not just accept the power of these tools, even if we do not understand fundamental reasons for the predicted performance? Yet we believe the complexity of these designs does not rival philosophical interpretations of quantum mechanics; we *should* aim to understand how they “truly” work. Such understanding might be exemplified by whether we can answer: what are the fundamental limits to the performance we are optimizing for? Why?

**Advances: Intelligent simulation and rational design**

Physically grounded, interpretable approaches can reveal *why* a design works and *how* it can be improved. Between brute-force simulations involving billions of variables and overly simplistic models that, while insightful, lack predictive power, there exists an intermediate paradigm we call *intelligent simulation*. This approach seeks to combine the strengths of both extremes. Ideally, an intelligent simulation identifies a small number of key latent parameters, relates them to physically meaningful quantities for direct interpretability, and still leverages the high dimensionality of the underlying system to achieve accurate predictions. We illustrate its power through three examples.

The first example concerns the development of grating theory in the 1970s. At that time, limited computational resources necessitated the design of highly efficient simulation methods, often grounded in sophisticated mathematical formulations with rapidly converging expansions. For instance, Rayleigh expansions were systematically employed in both the substrate and superstrate to enforce outgoing wave conditions, enabling a clear analytical distinction between propagating and evanescent waves. By contrast, most modern simulations that rely on finite-difference time-domain or finite-element methods, typically enforce these conditions with perfectly matched layers. While powerful and general, these layers tend to obscure physical interpretation and are often computationally demanding.

Furthermore, key physical quantities—such as the poles and zeros of the scattering operator—were frequently computed explicitly. Dispersion relations of bound states in the continuum were explicitly characterized, either as complex-valued frequencies as functions of the real part of the wavevector, or conversely as complex wavevectors as functions of the real frequency. Zeros, in particular, were often exploited to design unitary scattering efficiencies in selected diffraction orders, see, e.g., [3]. Today, such explicit characterization is seldom carried out, even though the set of poles—the spectrum of the associated non-Hermitian operator—encodes the most fundamental information about the system.

The second example is more recent and dates from a period when finite-difference time-domain and finite-element methods began to dominate computational electromagnetism. It concerns high-$Q$ photonic-crystal cavities formed by introducing defects (i.e., missing holes) in dielectric slabs. In the race towards high-$Q$, it was rapidly realized that cavity $Q$'s could be significantly enhanced by adjusting the position and size of the holes surrounding the defect. An emblematic illustration is the tenfold $Q$-enhancement with a three-hole defect cavity, achieved by shifting the two holes at the cavity extremities [4] (see Fig. 2).

The cavity mode can be understood primarily as a standing wave resulting from the interference of counterpropagating guided modes of a photonic-crystal waveguide with a single missing row. As such, it contains in-plane spatial frequency components exceeding the free-space wavenumber in the surrounding air cladding. Based on this observation, the authors of [4] introduced the concept of *gentle confinement*, whereby the cavity boundaries are engineered to avoid abrupt spatial variations of the mode profile. Instead, a gradual transition is imposed, ensuring that the mode remains predominantly composed of in-plane wavevectors lying below the air light line, thereby reducing radiation losses. While this concept was validated through "black-box" FDTD simulations, it does not directly reveal the physical mechanisms responsible for the observed enhancement: high-$Q$ cavities have all a reduced weight of wavevectors above the air light line and vice versa.

A complementary, more "modal" approach provides clearer physical insight. By modelling the cavity as a Fabry–Pérot resonator and using dedicated Bloch-mode numerical tools, the tenfold $Q$-enhancement is traced back to two key effects [5]. First, the group velocity of the photonic-crystal-waveguide guided mode is reduced by approximately a factor of two, effectively doubling the energy storage time within the cavity. Second, the shifted holes act as an impedance-matching section, adiabatically transforming the propagating Bloch mode of the waveguide into the evanescent Bloch mode of the surrounding photonic-crystal mirrors (Fig. 1) and effectively reducing out-of-plane scattering losses at the waveguide terminations.

Identifying these mechanisms provides valuable and transferable design principles. For instance, $Q$-factor enhancement through mode slowdown can be systematically exploited in a variety of geometries, including ring resonators and plasmonic nanoresonators. Similarly, impedance matching via gradual Bloch-mode tapering can be exploited across a wide range of photonic structures, including nanobeam cavities, micropillar Bragg cavities, and double-heterostructure PhC cavities [6].

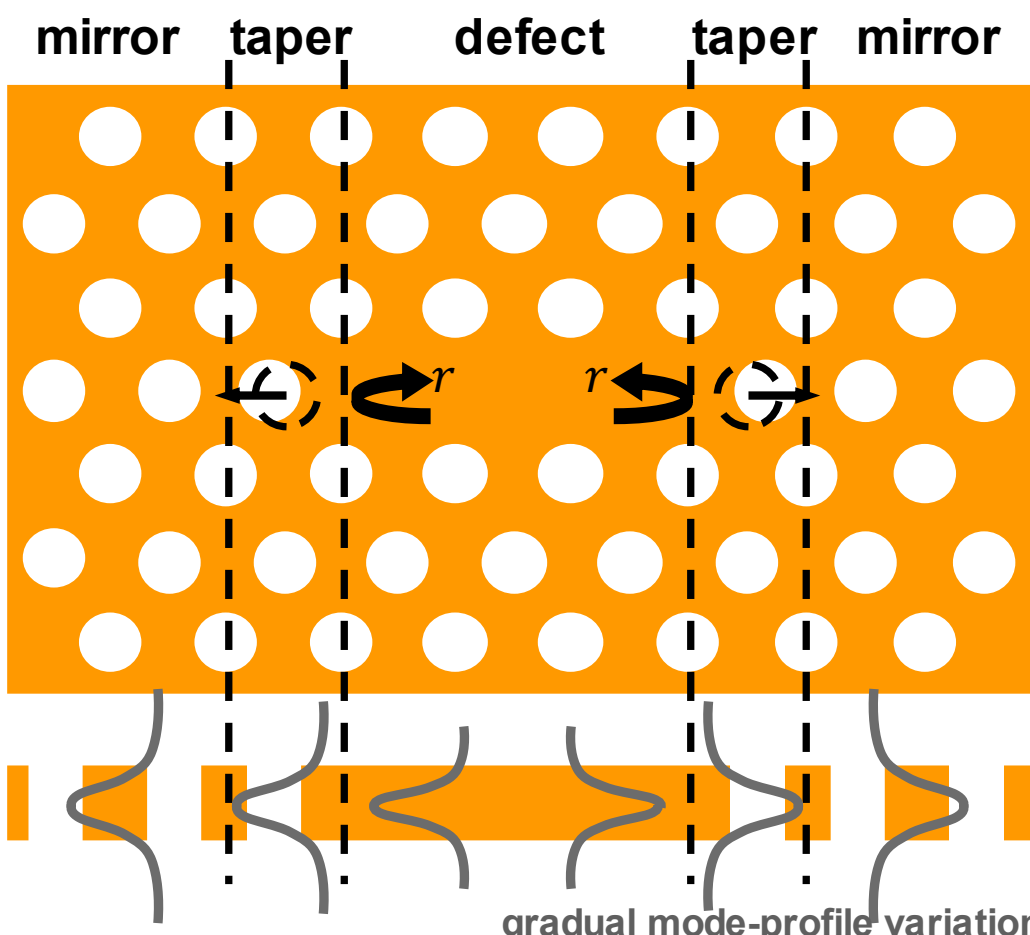

**Figure 2.** With a modal interpretation of gentle confinement with shifted holes, the Q-factor enhancement results from the slowdown of the guided Bloch mode in the cavity defect and its progressive impedance adaptation to the evanescent Bloch modes of the mirrors.

A third example arose recently in the design of a scattering device for augmented-reality applications [7]. Ideally, the device scatters strongly around a single wavelength (e.g., the source wavelength in the display), for many angles, while otherwise transparent across a wide range of frequencies and angles (e.g., radiation from the outside world, passing through the device). In a first attempt, the authors applied topology optimization to pixelated metasurfaces. They found designs that could concentrate their scattering within a relative bandwidth of $\Delta\omega/\omega = 1/30$ (far narrower than plasmonic approaches), but only for a small range of incoming angles, $\Delta\theta = \pm 12°$. The authors then tried a rational design, inspired by the flat bands of Moiré systems, and discovered simple two-parameter structures with guided-mode resonances that fit within the same $1/30$ bandwidth but with a much broader angular range of $\pm 45°$.

Could such simple designs really outwit computational inverse design? Not strictly: the background scattering in the latter designs was increased significantly. As measured by the numerical objective used in the optimization, the latter designs are worse. However, rational analysis revealed that the background scattering for these structures appear Fabry-Perot-like, motivating the introduction of an additional thin-film layer separately optimized as an anti-reflection coating for these structures. With this coating, the rational designs outperform the computationally optimized structures on every dimension. Physical insight reshaped the design "architecture" in a critically important way.

These examples illustrate a common pattern: understanding the latent physics of a problem—whether through modal decomposition, spectral characterization, or symmetry-inspired design—yields not only better-performing devices but also transferable principles applicable across geometries and physical systems. More generally, photonic design problems typically involve a multitude of hyperparameters that influence their "optimal" solutions, which are difficult to identify without physical insight. A natural question then arises: can these low-dimensional descriptors be discovered automatically, and can neural networks be trained to express their knowledge in physically interpretable form?

## Conclusions

Foundation models are not the culmination of AI, but merely its first operational layer. A second layer, called wrappers is now emerging to mediate, extend, and increasingly control the use of these models. In nanophotonics, this evolution is already visible in the rise of agentic frameworks [8-9]. These systems exhibit autonomy, retain memory, generate code, embed numerical Maxwell solvers, and orchestrate complex, multi-step workflows. Rather than simply executing tasks, they decompose problems, construct plans, and iteratively refine solutions—blurring the boundary between tool and collaborator. Yet, despite these advances, the collaboration remains fundamentally superficial. Today's systems excel at optimization but fall short at understanding. Their internal reasoning is largely inaccessible, leaving users to trust outputs without insight.

If AI is to become a true scientific partner, it must move beyond performance and toward interpretability. This requires shedding light on the *latent knowledge* encoded in hidden layers by making explicit which features drive decisions and why [10]. More radically, it calls for training neural networks to compress their knowledge into a small number of physically meaningful, low-dimensional parameters that can be directly understood by scientists. Without this step, AI risks remaining a powerful but opaque optimizer—efficient, yet epistemically limited. Similarly, this is a call to researchers: make the effort to understand your design. Ask your neural network to explain it, and strive for real insight. Only then will a genuine dialogue emerge between you and the neural network, which may lead to discover unexpected phenomena, such as a thin metal film acting as a lens with super-resolution capabilities.

Looking further ahead, digital scientists may recursively take over each step of this hierarchy, proposing and carry out their own research agenda, promising to drive an exponential acceleration in the growth of knowledge. Following Geoffrey Hinton's suggestion in his Nobel Prize remarks, "AI will enable us to create highly intelligent and knowledgeable assistants who will increase productivity in almost all industries."

However, it is not yet today that humanoid researchers orchestrate several $b^n$ digital scientists for a flurry or new discoveries. We can turn to chess for a relevant analogy. Even after "Deep Blue" defeated Gary Kasparov, the greatest "chess player" in the world was not in fact a computer program. For the next ten or fifteen years, the best chess was played by humans working in tandem with computers, each leveraging its own advantages. We might anticipate similar progress in fields such as photonics: leveraging cutting-edge computational tools while providing our own human advantages ("understanding," "intuition," etc.) may yield the greatest return. At least for a few years.

## Acknowledgements

PL would like to thank Pierre Chavel for many invaluable insights and constructive critiques, which have spanned over four decades. The authors also thank Amaury Badon and Louis Forestier for fruitful comments.